\documentclass[
    amsmath,
    amssymb,
    aps,
    twocolumn,
    superscriptaddress
]{revtex4-2}

\usepackage{graphicx}
\usepackage{amsmath}

\usepackage{color}

\newcommand{\position}{\mathbf{r}}
\newcommand{\cell}{\mathbf{n}}
\newcommand{\latticevector}{\boldsymbol{\ell}}
\newcommand{\height}{h}
\newcommand{\radius}{R}
\newcommand{\modulation}{\eta}
\newcommand{\displacement}{\mathbf{U}}
\newcommand{\rotation}{\mathbf{\boldsymbol\theta}}
\newcommand{\Rotation}{\mathbf{\boldsymbol\Theta}}
\newcommand{\stress}{\mathbf{\sigma}}
\newcommand{\strain}{\mathbf{\epsilon}}
\newcommand{\extension}{\epsilon}
\newcommand{\edgedir}{\hat{r}}

\newcommand{\Extensions}{\boldsymbol{\epsilon}}
\newcommand{\Displacements}{\mathbf{U}}
\newcommand{\compatibility}{\mathbf{B}}

\newcommand{\wavenumber}{\xi}
\newcommand{\wavevector}{\boldsymbol{\wavenumber}}
\newcommand{\dynamical}{\mathbf{K}}
\newcommand{\frequency}{\omega}

\newcommand{\mass}{\mathbf{M}}

\newcommand{\decay}{\kappa}
\newcommand{\winding}{\nu}
\newcommand{\oop}{\boldsymbol{\delta}}
\newcommand{\youngsmod}{E}
\newcommand{\momOfI}{I}
\newcommand{\shearCorr}{\mu}
\newcommand{\shearMod}{G}
\newcommand{\polarMom}{J}

\newcommand{\response}[1]{{\color{black}#1}}

\begin{document}

\title{Omnimodal topological polarization of bilayer networks: analysis in the Maxwell limit and experiments on a 3D-printed prototype}

\author{Mohammad Charara}
\thanks{Authors share equal contribution.}
\affiliation{Department of Civil, Environmental, and Geo- Engineering, University of Minnesota, Minneapolis, MN 55455, USA}
\author{James McInerney}
\email{jmcinern@umich.edu}
\thanks{Authors share equal contribution.}
\author{Kai Sun}
\author{Xiaoming Mao}
\email{maox@umich.edu}
\affiliation{Department of Physics, University of Michigan, Ann Arbor, MI 48109, USA}
\author{Stefano Gonella}
\email{sgonella@umn.edu}
\affiliation{Department of Civil, Environmental, and Geo- Engineering, University of Minnesota, Minneapolis, MN 55455, USA}
\date{\today}

\begin{abstract}\response{
    Periodic networks on the verge of mechanical instability, called Maxwell lattices, are known to exhibit zero-frequency modes localized to their boundaries. Topologically polarized Maxwell lattices, in particular, focus these zero modes to one of their boundaries in a manner that is protected against disorder by the reciprocal-space topology of the lattice's band structure. Here, we introduce a class of mechanical bilayers as a model system for designing topologically protected edge modes that couple in-plane dilational and shearing modes to out-of-plane flexural modes, a paradigm that we refer to as \emph{omnimodal polarization}. While these structures exhibit a high-dimensional design space that makes it difficult to predict the topological polarization of generic geometries, we are able to identify a family of mirror-symmetric bilayers that inherit the in-plane modal localization of their constitutive monolayers whose topological polarization can be determined analytically. Importantly, the coupling between the layers results in the emergence of omnimodal polarization, whereby in-plane and out-of-plane edge modes localize on the same edge. We demonstrate these theoretical results by fabricating a mirror-symmetric, topologically polarized kagome bilayer consisting of a network of elastic beams via additive manufacturing and confirm this finite-frequency polarization via finite element analysis and laser-vibrometry experiments. 
    }
\end{abstract}

\maketitle

\section{Introduction}
Topological phases of matter have properties that are characterized by a topological invariant, in contrast to conventional phases of matter that are characterized by symmetry breaking. \response{Importantly, these topological phases are defined by the reciprocal-space band structure of the lattice and are generically immune to disorder. For example, certain metallic compounds have bulk band structures that classify them as topological insulators possessing conducting states localized to their boundary~\cite{hasan2010colloquium,qi2011topological}. Similarly, a variety of mechanical systems can exhibit topologically protected, one-way elastic waves~\cite{prodan2009topological,susstrunk2015observation,wang2015topological,nash2015topological,wang2015coriolis,ma2019valley,torrent2013elastic,mousavi2015topologically,vila2017observation,miniaci2018experimental,chaunsali2018subwavelength,pal2017edge,miniaci2019valley}, which can be exploited for vibration isolation or waveguiding in mechanical metamaterials~\cite{bertoldi2017flexible,xin2020topological}.}

\response{
A special class of topological mechanical metamaterials, called \emph{Maxwell lattices}, have equal numbers of constraints and degrees of freedom, which places them near the onset of the rigidity transition~\cite{maxwell1864calculation,calladine1978buckminster,jacobs1995generic}. Interestingly, such Maxwell lattices can exhibit zero-frequency modes that localize to their boundaries~\cite{kane2014topological,lubensky2015phonons,mao2018maxwell,rocklin2017directional,rocklin2017transformable,stenull2016topological,zhou2018topological,socolar2017mechanical,baardink2018localizing}, e.g., edges in two dimensions or surfaces in three dimensions. The relative number of such zero modes on opposite sides of the lattice specifies the \emph{topological polarization} of the structure and distinguishes between distinct topological phases. The main consequence of this topological polarization is the ability of the lattice to display an excess of softness on a particular edge without requiring generalized softness of the entire system, thereby preserving load-bearing capabilities. The fact that the polarization is topological implies that the phenomenon is not a property specific to the edge, and therefore subjected to the variability of the edge properties; rather it is a property of the bulk that manifests at the edges (i.e., the ``bulk-edge correspondence''), and therefore it is guaranteed to persist as long as the topological properties of the bulk remain unchanged. Moreover, these topologically polarized Maxwell lattices possess a number of additional notable properties including transformability~\cite{guest2003determinacy,borcea2010periodic}, mechanical solitons~\cite{chen2014nonlinear,zhou2017kink}, mechanical Weyl modes~\cite{rocklin2016mechanical,stenull2016topological,socolar2017mechanical,baardink2018localizing}, stress focusing~\cite{paulose2015selective}, and robustness against fracture~\cite{zhang2018fracturing}.
}

\response{
The experimental realization of these topologically polarized Maxwell lattices remains a challenge. The existence of non-trivial zero modes requires finely-tuned joints that eliminate friction and bending stiffness, which can be realized using pin-jointed truss structures for planar configurations~\cite{chen2014nonlinear,rocklin2017transformable}. In contrast, \emph{structural} realizations, such as those fabricated via additive manufacturing, feature structural elements whose kinematics exhibit additional interactions that over-constrain Maxwell geometries. While this implies that such structural Maxwell lattices cannot possess zero modes, the boundary modes predicted for ideal Maxwell lattices migrate to finite frequencies. For example, Ref.~\cite{ma2018edge} investigates a topologically polarized, structural kagome lattice realized via waterjet cutting of a polymer sheet and experimentally demonstrates a combined edge- and frequency-selectivity of the mechanical response. Importantly, heuristic theoretical models show that this polarization can vanish when the joints become too stiff, thereby leading to bulk wave propagation at all frequencies~\cite{stenull2019signatures}.
}

Moreover, although the in-plane, topologically polarized modes of two-dimensional Maxwell lattices are well understood, fundamental questions remain about the mechanics of two-dimensional Maxwell lattices embedded in three-dimensional space. This is the scenario of thin plate-like structures, in which flexural modes dominate the low-energy mechanical response due to the relative scaling of their bending and stretching moduli~\cite{landau1986theory}. This feature can be manipulated for the deployment and transformation of mechanical structures by programming particular flexural modes into the sheet, such as the rigid folding modes of origami metamaterials~\cite{santangelo2017extreme,santangelo2020theory,dieleman2020jigsaw}. However, triangulated origami, which lies at the Maxwell point, possesses a hidden symmetry that pairs edge modes on opposite sides of the sheet, thereby preventing its topological polarization~\cite{mcinerney2020hidden}. While this hidden symmetry can be broken by introducing an equal number of quadrilateral faces and holes~\cite{finbow2012rigidity}, thereby yielding topologically polarized Maxwell kirigami~\cite{chen2016topological}, these quadrilateral faces exhibit additional low-energy modes, due to bending of the faces~\cite{schenk2011origami,filipov2017bar}, which are not necessarily localized. \response{Furthermore, such topologically polarized Maxwell kirigami sheets are challenging to realize experimentally and their large design space makes it difficult to characterize polarized geometries analytically. Thus, the state of the art suggests that the simultaneous topological polarization of in-plane and out-of-plane edge modes remains an important question, both theoretically and experimentally.

The mechanical bilayers introduced in Ref.~\cite{charara2021topological} present one possible solution. Such bilayers are designed heuristically to have one topologically polarized layer and one unpolarized layer. This polarization mismatch presents an incompatibility between the in-plane effective stiffness of the two layers so that the in-plane actuation of the edge modes couples to out-of-plane flexural deformations that localize at the same edge. However, this approach presents a few major limitations. First, the polarization of flexural waves requires activation of the in-plane strains through a force applied directly in the plane of the topological layer, while out-of-plane actuation leads strictly to bulk wave propagation, thus limiting the functional versatility of this configuration as a wave control device. Second, because this bilayer is constructed by attaching the two layers as continua, the system is not close to the Maxwell connectivity, and thus the modes of this bilayer are not characterized by a topological invariant, implying that the observed polarization is generally weak and does not feature any topological protection.

The present manuscript addresses the topological polarization of Maxwell bilayer geometries both in theory and in experiment. We design a bilayer lattice that exactly satisfies the Maxwell criterion when modeled as point masses connected by harmonic springs, permitting the rigorous definition of a topological index.  Furthermore, we choose a subset of Maxwell bilayer geometries that are characterized by a mirror symmetry between the top and bottom layers along with a particular choice of interlayer connections. We show that these two features endow the bilayer with simultaneous in-plane and out-of-plane edge modes whose localization is protected by a topological index, and is inherited from the planar geometry of the individual layers, a condition we refer to as \emph{omnimodal polarization}. We then consider the polarization of a lattice with the same geometry but composed of elastic beams where boundary conditions at the joints become incompatible with the zero modes without bending the beams, and show via finite element analysis that the polarized edge modes persist at finite frequencies. Finally, we corroborate these predictions through laser-vibrometry experiments on a 3D-printed prototype. 
}

\section{Results and Discussion}
\subsection{Topological Polarization of Maxwell Bilayers}
\response{
This section formally introduces the geometry of mirror-symmetric kagome bilayers and explores their topological polarization in the idealized spring-mass model. Importantly, we distinguish between \emph{coplanar} bilayers, where the vertices of a single layer lie in a plane, and \emph{non-coplanar} bilayers, where the vertices of each single layer lie at different heights. We show that the localized out-of-plane response of coplanar bilayers can be characterized analytically by the antisymmetric activation of the topological edge modes of their constituent monolayers; however, these coplanar bilayers also possess bulk zero modes that compromise the mode selection and topological protection of their edge modes. In contrast, non-coplanar bilayers possess strongly localized edge modes in the short-wavelength regime and their bulk zero modes are restricted to zero-dimensional points so that the edge modes are topologically protected. These results indicate that the imposed mirror-symmetry effectively reduces the design space of kagome bilayers to achieve omnimodal topological polarization by leveraging the well-characterized topological polarization of planar kagome monolayers (see SI Appendix).
}
\begin{figure*}[t]
    \centering
    \includegraphics{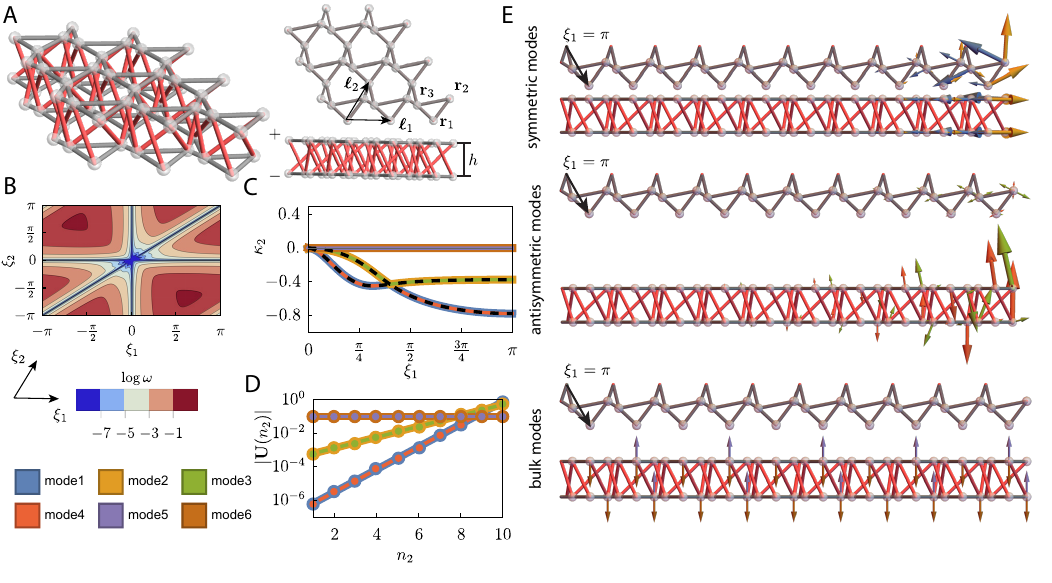}
    \caption{
    Coplanar, mirror-symmetric kagome bilayer. (A) An illustration of the lattice geometry with planar vertex positions, $\position_i$, lattice vectors, $\latticevector_{1,2}$, and layer separation $\height$. (B) Eigenfrequency, $\frequency$, for the lowest energy band over the two-dimensional Brillouin zone with wavenumbers, $\wavevector=(\wavenumber_1,\wavenumber_2)$, directed along the corresponding lattice vectors, revealing lines of bulk modes along wavevectors that exhibit the vertex-pair modes in Eqn.~(\ref{eq:pair}). (C) The decay rate of zero modes, $\decay_2$, along the $\latticevector_2$-direction as a function of the wavenumber, $\wavenumber_1$, in the $\latticevector_1$-direction. The dashed lines indicate the decay rates of the constitutive planar kagome lattices, at which the edge modes are doubly degenerate. (D) Norm of vertex displacements, $\lvert \Displacements(n_2) \rvert$ in each cell for the six zero modes with periodic boundary conditions $\wavenumber_1 = \pi$. (E) Illustration of the six zero modes in a supercell with periodic boundary conditions $\wavenumber_1 = \pi$ in the $\latticevector_1$-direction, where the size of the arrows denote the relative displacements of the vertices.
    }
    \label{fig:fig1}
\end{figure*}

\begin{figure*}[t]
    \centering
    \includegraphics{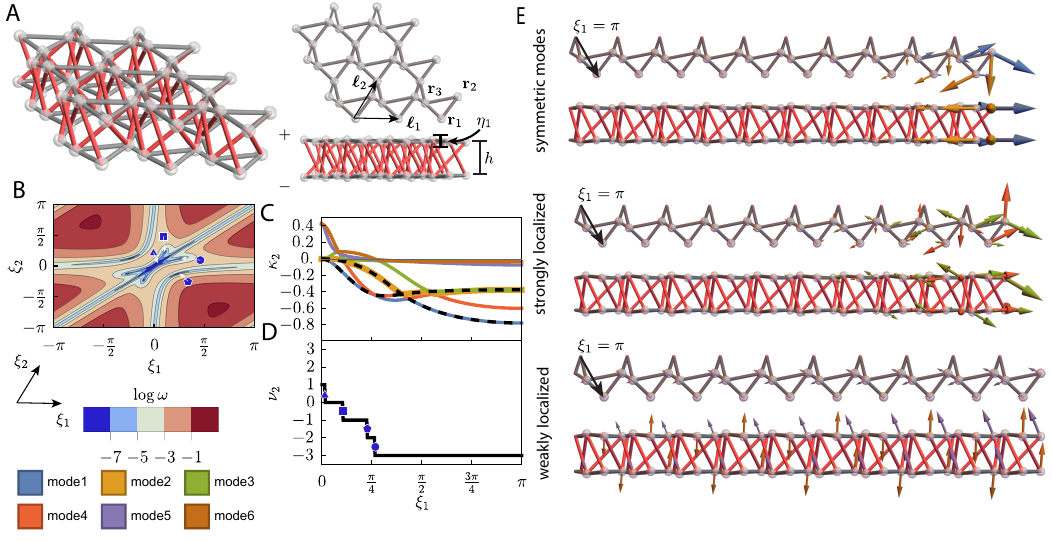}
    \caption{Non-coplanar, mirror-symmetric kagome bilayer. (A) An illustration of the lattice geometry with planar vertex positions, $\position_i$, lattice vectors, $\latticevector_{1,2}$, layer separation $\height$, and height modulation, $\modulation_i$. (B) Eigenfrequency, $\frequency$, for the lowest energy band over the two-dimensional Brillouin zone with wavenumbers, $\wavevector=(\wavenumber_1,\wavenumber_2)$, directed along the corresponding lattice vectors, revealing bulk modes at isolated wavevectors that denote Weyl points where one of the decay rates changes sign and the winding number is discontinuous. (C) The decay rate of zero modes, $\decay_2$, along the $\latticevector_2$-direction as a function of the wavenumber, $\wavenumber_1$, in the $\latticevector_1$-direction. The dashed lines indicate the decay rates of the constitutive planar kagome lattices. (D) The winding number, $\winding_2$, computed by integrating over the wavenumber, $\wavenumber_2$, in the $\latticevector_2$-direction from $-\pi$ to $\pi$ as a function of the wavenumber, $\wavenumber_1$, in the $\latticevector_1$-direction. (E) Illustration of the six zero modes in a supercell with periodic boundary conditions $\wavenumber_1 = \pi$ in the $\latticevector_1$-direction, where the size of the arrows denote the relative displacements of the vertices.}
    \label{fig:fig2}
\end{figure*}

\response{
Figs.~\ref{fig:fig1}A and~\ref{fig:fig2}A show examples of coplanar and non-coplanar, mirror-symmetric kagome bilayers, respectively. Both are composed of unit cells that contain six vertices, specified by the in-plane positions $\position_1 = (0, 0, 0)$, $\position_2 = (1, 0.3, 0)$, and $\position_3 = (0.5, \sqrt{3}/2 + 0.6, 0)$ and the layer separation $h = 1.5$ as: $\position^\pm_i = \position_i \pm (h/2 + \modulation_i) \hat{z}$, where $+(-)$ specifies a vertex in the upper (lower) layer. The vertex-dependent height modulation, $\modulation_i$, distinguishes the coplanar geometry, for which $\modulation_i = 0$, from the non-coplanar geometry, for which we choose $\modulation_1 = 0.1$ and $\modulation_2 = \modulation_3 = 0$. Since each vertex has three spatial degrees of freedom, six interlayer bonds must be introduced for the bilayers to achieve Maxwell coordination in the bulk and eliminate trivial zero modes. We utilize the particular choice which creates three pairs of adjoined vertices out of the six vertices in each unit cell \big($\position_2^-$ connects to $\position_1^+$ in cells $(0,0)$ and $(1,0)$, $\position_3^-$ connects to $\position_2^+$ in cells $(0,0)$ and $(-1,1)$, and $\position_1^-$ connects to $\position_3^+$ in cells $(0,0)$ and $(0,-1)$\big) which enables greater control over the topological polarization in contrast to alternative choices.} Each cell is related to its neighbors by the lattice vectors, $\latticevector_1 = (2,0,0)$ and $\latticevector_2 = (1, \sqrt{3},0)$, so that the position of vertex $i$ in cell $\cell = (n_1,n_2)$ is: $\position^\pm_i(\cell) = \position^\pm_i + n_1 \latticevector_1 + n_2 \latticevector_2$. Note that the analysis in this section is scale-independent so that the lengths are written in dimensionless units.

\response{In the spring-mass model,} the mechanical response of such bilayers is characterized by their normal modes. \response{This model considers each vertex to have a mass $m$ and each edge to be a spring with stiffness $k$.} The infinitesimal displacements, \response{$\displacement_i = (u_i, v_i, w_i)$ and  $\displacement_j = (u_j, v_j, w_j)$}, of vertices $i$ and $j$, respectively, extend the spring adjoining these vertices to first-order by an amount $\extension_{ij} = \edgedir_{ij} \cdot (\displacement_j - \displacement_i)$, thereby leading to a restoring force directed along their shared edge, $\mathbf{f}_{ij} = - k \extension_{ij} \edgedir_{ij}$, which causes the masses to oscillate about their equilibrium position with frequency \response{$\frequency=\sqrt{k/m}$}. Thus, these dynamical quantities are non-dimensionalized by introducing the timescale $\sqrt{m/k}$, and hence we take \response{$k = m = 1$}. The linear operator, \response{$\compatibility$}, called the \emph{compatibility matrix}, maps the vector of all displacements, \response{$\Displacements = (\displacement_1, \displacement_2, \ldots)$}, to the vector of all extensions, \response{$\Extensions = (\extension_{ij}, \extension_{ik}, \ldots)$}, so that such oscillations correspond to the eigenmodes, $\dynamical \Displacements = \frequency^2 \Displacements$, of the stiffness matrix, \response{$\dynamical = \compatibility^T \compatibility$ (see SI Appendix for remarks on notation)}. \response{Since the bilayers are periodic, their} normal modes are Bloch-periodic in the bulk. Thus, the compatibility matrix can be diagonalized into blocks \response{in reciprocal space} $\compatibility(\wavevector)$ for each wavevector $\wavevector = (\wavenumber_1,\wavenumber_2)$, \response{where the wavenumbers $\wavenumber_{1,2}$ are defined along the corresponding lattice vector $\latticevector_{1,2}$ (see Materials and Methods)}. The corresponding displacements vary between cells as $\Displacements(\cell) = \Displacements e^{i \wavevector \cdot \cell}$, where $\Displacements$ is an eigenvector of the Bloch-periodic \response{stiffness} matrix $\dynamical(\wavevector) = \compatibility(\wavevector)^\dagger \compatibility(\wavevector)$, with $\dagger$ denoting Hermitian conjugation. \response{The Maxwell coordination implies that one zero mode arises for each bond cut from the boundary in a finite-sized realization. Generically, these zero modes localize to an edge of the bilayer and are characterized by complex-valued wavenumbers where the imaginary part, $\decay = \textrm{Im} \wavenumber$, specifies the decay rate. Such edge modes can be realized in a \emph{supercell} with periodic boundary conditions in the $\latticevector_1$-direction and a finite number of cells with open boundary conditions in the $\latticevector_2$-direction. The vertex displacements then grow (decay) across the supercell as $e^{-\decay_2 n_2}$ for $\decay_2 < 0~(\decay_2 > 0)$.} The topological aspect of these modes arises from a bulk-boundary correspondence whereby the topology of the Brillouin zone dictates the localization of edge modes. \response{In the absence of bulk zero modes,} this is quantified by an integer-valued topological invariant, called the \emph{winding number}

\begin{equation} \label{eq:winding}
    \winding_2(\wavenumber_1) = \frac{1}{2 \pi} \int_{-\pi}^{+\pi} d\wavenumber_2 \frac{\partial}{\partial_{\wavenumber_2}} \text{Im} \log \det \compatibility(\wavenumber_1,\wavenumber_2),
\end{equation}

\noindent
which determines the relative number of zero modes localized to each edge (see Materials and Methods). \response{This type of winding number was introduced in Ref.~\cite{kane2014topological} for in-plane topological floppy modes.}

The compatibility matrix for generic bilayers maps in-plane ($\parallel$) and out-of-plane ($\perp$) degrees of freedom in the upper ($+$) and lower ($-$) layers to constraints on the upper ($+$), lower ($-$), and interlayer ($\pm$) bonds:

\begin{equation} \label{eq:compatibility}
    \begin{pmatrix}
        \Extensions^- \\
        \Extensions^+ \\
        \Extensions^\pm
    \end{pmatrix}
    =
    \begin{pmatrix} 
        \compatibility_{\parallel}^- & \mathbf{0} & \compatibility_{\perp}^- & \mathbf{0} \\
        \mathbf{0} & \compatibility_{\parallel}^+ & \mathbf{0} & \compatibility_{\perp}^+ \\
        \multicolumn{2}{c}{\compatibility_{\parallel}^\pm} & \multicolumn{2}{c}{\compatibility_{\perp}^\pm}
    \end{pmatrix}
    \begin{pmatrix}
        \Displacements^-_\parallel \\
        \Displacements^+_\parallel \\
        \Displacements^-_\perp \\
        \Displacements^+_\perp
    \end{pmatrix}.
\end{equation}

\noindent 
Importantly, the mirror symmetry implies that the intralayer, in-plane constraints are equal for each layer, $\compatibility_\parallel^- = \compatibility_{\parallel}^+$, so that \response{the normal modes of the constitutive planar monolayers are normal modes of the bilayer with strictly in-plane displacements. This means that the topologically polarized zero modes of the planar monolayers are inherited by both the coplanar and non-coplanar mirror-symmetric bilayers as symmetric modes that displace vertices in the top and bottom layers identically:}

\begin{equation} \label{eq:symmetric} 
    \Displacements_{\text{sym}} = (+\Displacements_\parallel, +\Displacements_\parallel, \mathbf{0}, \mathbf{0}).
\end{equation}

\noindent
\response{The character of these symmetric modes for the coplanar (non-coplanar) bilayer is shown by the decay rates in Fig.~\ref{fig:fig1}C~(\ref{fig:fig2}C), which overlap with the black dashed lines corresponding to the decay rates of the planar monolayer, and the two strictly in-plane mode shapes in Fig.~\ref{fig:fig1}E~(\ref{fig:fig2}F), where the size of the arrows indicate the relative displacements. Since the mirror-symmetric kagome bilayer has six total intercellular bonds along the $\latticevector_2$-direction (two from each layer and another two adjoining the layers together), it can exhibit up to six zero modes localized to a single edge for any value of the transverse wavenumber $\wavenumber_1$.} The character of the four remaining zero modes depends on the magnitude of any height modulations and the assignment of interlayer connections.

For \emph{coplanar} bilayers, the intralayer, out-of-plane constraints are trivially satisfied, $\compatibility_\perp^- \Displacements_\perp^- = - \compatibility_\perp^+ \Displacements_\perp^+ = \mathbf{0}$, so that the out-of-plane displacements are only constrained by the interlayer connections. \response{This simplification allows the four remaining modes to be categorized as two types: (i) antisymmetric modes that couple in-plane and out-of-plane deformations of the bilayer and (ii) bulk modes that displace isolated sets of vertices (which depend on the choice of interlayer connections) out-of-plane. The antisymmetric modes are a straightforward generalization of the symmetric modes where the zero modes of the planar monolayers are activated antisymmetrically between the two layers, which generates incompatibility that can only be resolved by simultaneously displacing the vertices out-of-plane as illustrated by the two antisymmetric mode shapes in Fig.~\ref{fig:fig1}E:}

\begin{equation} \label{eq:antisymmetric}
    \Displacements_{\text{asym}} = (+\Displacements_\parallel, -\Displacements_\parallel, \Displacements_\perp^-, \Displacements_\perp^+),
\end{equation}

\noindent
where $\Displacements_\perp^+ \neq \Displacements_\perp^-$ due to asymmetry of the interlayer connections. \response{The flexural nature of these antisymmetric modes can be understood in analogy to the bending of an elastic plate, where the extension of the top of the plate is accompanied by contraction of the bottom of the plate. The key aspect in the mirror-symmetric bilayer is that this deformation localizes according to the topological polarization of the planar monolayers, thus the corresponding decay rates are doubly-degenerate as shown in Figs.~\ref{fig:fig1}C.  Since the interlayer connections only couple isolated pairs of vertices in the unit cell, which connect to one another along rows across the whole lattice, the coplanar bilayer possesses zero modes that displace these pairs of vertices strictly out-of-plane}

\begin{equation} \label{eq:pair}
    \Displacements_{\text{pair}} = ( \mathbf{0}, \mathbf{0}, \oop_i^-, \oop_j^+ ),
\end{equation}

\noindent
where $\oop_i^- = \oop_j^+$ are the isolated out-of plane displacements of vertices $i$ and $j$ in the lower and upper layers respectively, which are connected via interlayer bonds, as illustrated by the two strictly out-of-plane mode shapes in Fig.~\ref{fig:fig1}E. These vertex-pair modes exist at wavevectors \response{perpendicular to the lines formed by these vertex pairs, so that the whole line displaces out of plane by the same amount, stretching no bonds}: $\oop_i(\cell') = \oop_i e^{\wavevector \cdot \cell'} = \oop_i$. The vertex-pair modes in Eqn.~(\ref{eq:pair}) are given by $\oop_3^- = \oop_1^+$ for $\wavenumber_2 = 0$, $\oop_2^- = \oop_3^+$ for $\wavenumber_2 = \wavenumber_1$, and $\oop_1^- = \oop_2^+$ at $\wavenumber_1 = 0$, as shown by the lines of bulk modes apparent in the vanishing eigenfrequencies of the lowest band shown in Fig.~\ref{fig:fig1}B, and the doubly-degenerate lines of $\decay_2 = 0$ at $\wavenumber_1 \neq 0$ shown in Fig.~\ref{fig:fig1}C. Note that alternative choices of interlayer connections can change the character of these trivial modes, such as the connections which exhibit a flat band discussed in the SI Appendix. Importantly, the vertex-pair modes are bulk zero modes that inhibit the calculation of the winding number, \response{so that the four edge modes (the symmetric and antisymmetric modes) cannot be explicitly shown to be topologically robust, and make it difficult to selectively actuate the edge modes.}

\response{For the \emph{non-coplanar} bilayer, the four zero modes do not admit the simple analytical descriptions that they do for the coplanar bilayer. This is because} the height modulation, $\modulation_i$, breaks the geometric singularity that gives rise to these bulk modes and makes the intralayer edges constrain out-of-plane displacements, $\compatibility^+_\perp = - \compatibility^-_\perp \neq \mathbf{0}$. \response{This height modulation lifts the lines of bulk zero modes corresponding to the vertex pair modes in Eqn.~\ref{eq:pair} and introduces Weyl points, labeled by the blue markers in Fig.~\ref{fig:fig2}B, at the origin of the Brillouin Zone that move to larger wavenumbers as the magnitude of the modulation increases. Consequently, the antisymmetric modes in Eqn.~(\ref{eq:antisymmetric}) remain \emph{strongly} localized and the vertex-pair modes in Eqn.~(\ref{eq:pair}) become \emph{weakly} localized in the short wavelength regime, as indicated by their respective decay rates in Fig.~\ref{fig:fig2}C, even though the character of their modes shapes changes as shown in Fig.~\ref{fig:fig2}F: the strongly localized modes retain their out-of-plane character but no longer exhibit an antisymmetry between the top and bottom layers and the weakly localized modes acquire an in-plane component. The significance is that now, all six of the zero modes of the non-coplanar bilayer localize to the top edge at short wavelength and four of these zero modes exhibit both in-plane and out-of-plane character. As a result, the winding number is well-defined, as shown in Fig.~\ref{fig:fig2}D, away from the Weyl points where it undergoes discontinuities labeled by the blue markers that correspond to a mode moving to the opposite edge, thus yielding the desired omnimodal topological polarization. The exact location of these Weyl points is highly nontrivial to predict analytically from the geometry, which can lead to undesirable zero modes on the edge opposite of the polarization direction. Importantly, the present choice of interlayer connections limits such undesirable effects, whereas alternative choices can lead to more dramatic changes, including the presence of zero modes strongly localized to the opposite edge (see SI Appendix).}

The analysis of the topologically polarized, mirror-symmetric kagome bilayer presented here can be extended to bilayers composed of topological square lattices~\cite{rocklin2017transformable,rocklin2016mechanical}, bilayers without mirror symmetry which do not exhibit strictly in-plane modes, and bilayers with interlayer connections that yield flat bands in the coplanar limit (see SI Appendix). Thus, investigations of less restrictive geometries could provide greater control over the presence of bulk modes and exhibit more strongly polarized flexural modes. \response{However, the mirror-symmetric case provides an intuitive explanation for the existence of omnimodal edge modes in the coplanar limit and dramatically reduces the design space for topologically polarized, flexural mechanical metamaterials.}

\subsection{Edge Modes in Structural Bilayers: Supercell Analysis Simulations}

\begin{figure}[t!]
    \includegraphics{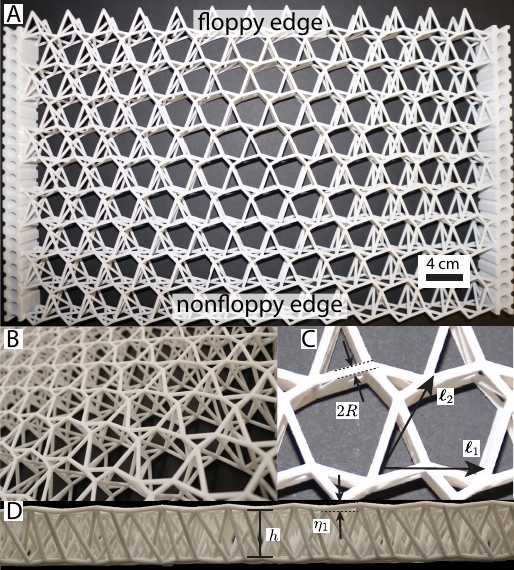}
    \caption{3D-printed structural lattice: (A) Top-down and (B) angled views of the symmetric kagome bilayer introduced in Fig.~\ref{fig:fig2} constructed via SLS and composed of glass-fiber-reinforced nylon beams with $\radius = 1.5$mm. The lattice geometry is proportional to that of the spring-mass model. (C) Close-up of the lattice unit cell with lattice vectors $\latticevector_1 = 2 (2, 0 ,0)~\textrm{cm}$ and $\latticevector_2 = 2 (1, \sqrt{3}, 0)~\textrm{cm}$. (D) Front view of the bilayer with layer height separation $\height = 3~\textrm{cm}$ and height modulation $\modulation_1 = 2~\textrm{mm}$.}
    \label{fig:system}
\end{figure}

\begin{figure*}[t]
	\centering
	\includegraphics{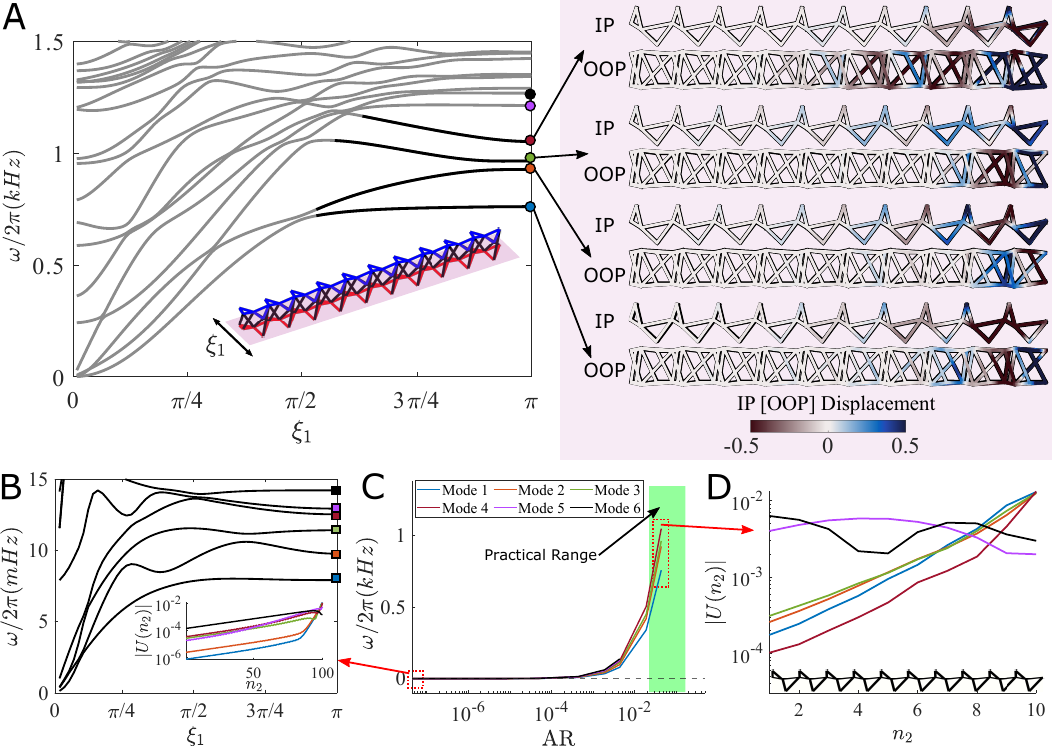} 
	\caption{(A) Band diagram of a 10-unit bilayer supercell (angled view shown in the inset), calculated using 3D Timoshenko beam finite element discretization; mode shape of modes 1-4 at $\wavenumber=\pi$, with in-plane (IP) and out-of-plane (OOP) deformation plotted with color intensity proportional to displacement along $\hat{y}$ and  $\hat{z}$, respectively (each IP and OOP mode shape is normalized by the highest displacement value of that data set) - here $\hat{y}$ is chosen to represent in-plane displacement due to it being normal to the top and bottom edges; (B) supercell band diagram for low aspect ratio (AR) beams, (approaching the spring-mass model limit) showing six decaying modes separated from the bulk band, whose decay rates at $\wavenumber=\pi$ - highlighted by the square markers - are shown in the displacement vs. cell index plot in the inset); (C) frequency of the six lowest modes as a function of AR (the legend here also applies to panel B and D); (D) Displacement vs. cell index for the first 6 modes of the supercell in panel A [with AR value compatible with manufacturing constraints] (highlighted be the circle markers), showing four decaying modes and two bulk modes. The green region in panel B highlights the practical region of AR - beam with ARs below are filaments outside the realization of most additive manufacturing methods, and those above intersect each other and become too thick to approximate with beam theory.}\label{fig:beam_bdms}
\end{figure*}

\response{
The spring-mass model presented in the previous section provides a tool for analytical treatment of the bilayer, and assessing its topological roots. In terms of realistic physical systems, this framework also captures the behavior of truss-based networks of rods connected by ideal joints, which are extremely difficult to fabricate. Here, we consider a realistic structural bilayer obtained via additive manufacturing, and therefore amenable to physical testing, more appropriately modeled as a network of elastic beams. Importantly, these beams~
meet at internal clamps which locally prevent their free relative rotation, in contrast to the perfect hinges in the spring-mass model for the \emph{ideal Maxwell} bilayer. As a result, the system becomes over-constrained, despite nominally maintaining the critical coordination number, and we cannot formally calculate a topological invariant. However, we elucidate the bilayer's ability to polarize in-plane and out-of-plane modes and, through a systematic parametric study, we can provide evidence that its behavior possesses \emph{topological origin}, rooted the topological polarization of its ideal Maxwell counterpart.}

\response{
To this end, we construct the $10\times13$ cell kagome bilayer prototype shown in Fig. \ref{fig:system}, featuring the geometry and connectivity of the non-coplanar bilayer in Fig. \ref{fig:fig2}, via selective laser sintering (SLS) with glass-fiber-reinforced nylon (see Materials and Methods).} In the structural lattice, the springs connecting the vertices are replaced with beams of radius $\radius=1.5\,\textrm{mm}$, whose mass is distributed along the length of the beam, rather than lumped at the vertices. This lattice features an interlayer height $h=3\,\textrm{cm}$ and a height modulation $\modulation_1=2\,\textrm{mm}$. \response{Note that solid flanges are fabricated on the left and right edges of the lattice to act as clamping sites for the experiments presented in the following section. }

\response{
We model the of the structural lattice as 3D Timoshenko beams~\cite{goodier1970theory, karadeniz2013finite} subject to clamped boundary conditions which preserve the relative angles between beams meeting at a joint and invoke bending of the beams (see Materials and Methods). More generally, Timoshenko beams are capable of undergoing axial, flexural, shear, and torsional deformations. Note that the choice of the Timoshenko model is inspired by a criterion of generality, and we obtain nearly identical results using the Euler-Bernoulli model, which omits shear (an effect that becomes more relevant for short and stubby beams).} \response{We discretize each bond via the finite element method (FEM) framework (see Materials and Methods), into multiple beam elements, whose nodes each feature three positional degrees of freedom and three cross sectional rotations (see SI Appendix).} We then construct a supercell by aggregating 10 cells in the $\latticevector_2$ direction, and applying free boundary conditions at the ends and Bloch-periodic boundary conditions with respect to wavenumber $\wavenumber_1$ in the $\latticevector_1$ direction (see Materials and Methods) to calculate the bilayer's band diagram. The resulting supercell band diagram is shown in Fig.~\ref{fig:beam_bdms}A. We seek the edge modes among the low-frequency modes that fall below the bulk band near the edge of the Brillouin zone, in analogy with the scenario described in Ref.~\cite{charara2021topological} (the relevant portion of the branches are plotted in black). The mode shapes of the lowest four modes at $\wavenumber_1 = \pi$ are shown where the in-plane deformation is shown with color intensity proportional to displacement along $\hat{y}$ (the Cartesian direction normal to the open boundaries of the supercell and lattice) and the out-of-plane deformation is shown with color proportional to displacement along $\hat{z}$. 

Both the in-plane and the out-of-plane deformation exhibit strong localization at the top end of the supercell, with a comparable decay rate. \response{This localization is quantified by the norm of the total nodal displacements in each cell, which is shown to increase exponentially towards the top edge of the supercell ($n_2=10$) in Fig. \ref{fig:beam_bdms}D}. The most striking observation is that all four modes below the bulk band are localized. This is a considerable qualitative improvement over the results presented in Ref.~\cite{charara2021topological}, in which only two polarized edge modes were found and, more importantly, where the overall signature of polarization in the flexural response was significantly milder. Moreover, the localization of these modes highlights that the top and bottom edges of this bilayer serve as the floppy and nonfloppy edges, respectively, consistent with the predictions from the ideal Maxwell lattice. 

\response{
Although four low-frequency modes localize in the structural lattice, the ideal Maxwell case predicts six total edge modes. Since these zero modes involve unimpeded rotations about the perfect hinges with no elongation of the springs, which is disallowed in the structural model due to the clamped boundary conditions at the beam ends, the beam's behavior should approach (albeit never reach) the behavior of the ideal Maxwell case when the cost of beam bending becomes negligible (to the cost of beam extension). This limit can be achieved by tuning the beams' aspect ratio ($AR=\radius/L$, where $L$ is the length of the shortest bond in the unit cell) because the axial stiffness scales with $\radius^2$ while the bending stiffness scales with $\radius^4$. Thus, as the $AR$ diminishes, the bending stiffness becomes minute and a greater amount of flexural deformation is allowed in the vicinity of the structural hinges. Indeed, in Fig.~\ref{fig:beam_bdms}C we compare the frequency of the topologically polarized modes as a function of $AR$, monitoring the frequencies of the first six modes (until they enter the bulk band and can no longer be identified), finding that as $AR$ decreases, the frequencies approach zero. The band diagram of the low $AR$ case (Fig.~\ref{fig:beam_bdms}B) shows that as we decrease $AR$, two additional modes drop from the bulk band and become localized, as confirmed by plotting the norm of the total displacements in each cell, shown in the inset. Note that a $100$ unit-cell supercell was used here due to the low decay rate exhibited by some of the modes, which require a large number of cells to appreciate. To confirm the contrast with the results applicable to our prototype, decay rates for a $100$ unit-cell supercell with the geometric characteristics (i.e., $AR$) of the prototype are shown in the SI Appendix, Fig.~S3; even with the longer supercell we still only observe four localized modes.
}

\subsection{Wave Propagation: Experiments and Full-Scale Simulations}

\begin{figure*}[t!]
	\centering
	\includegraphics{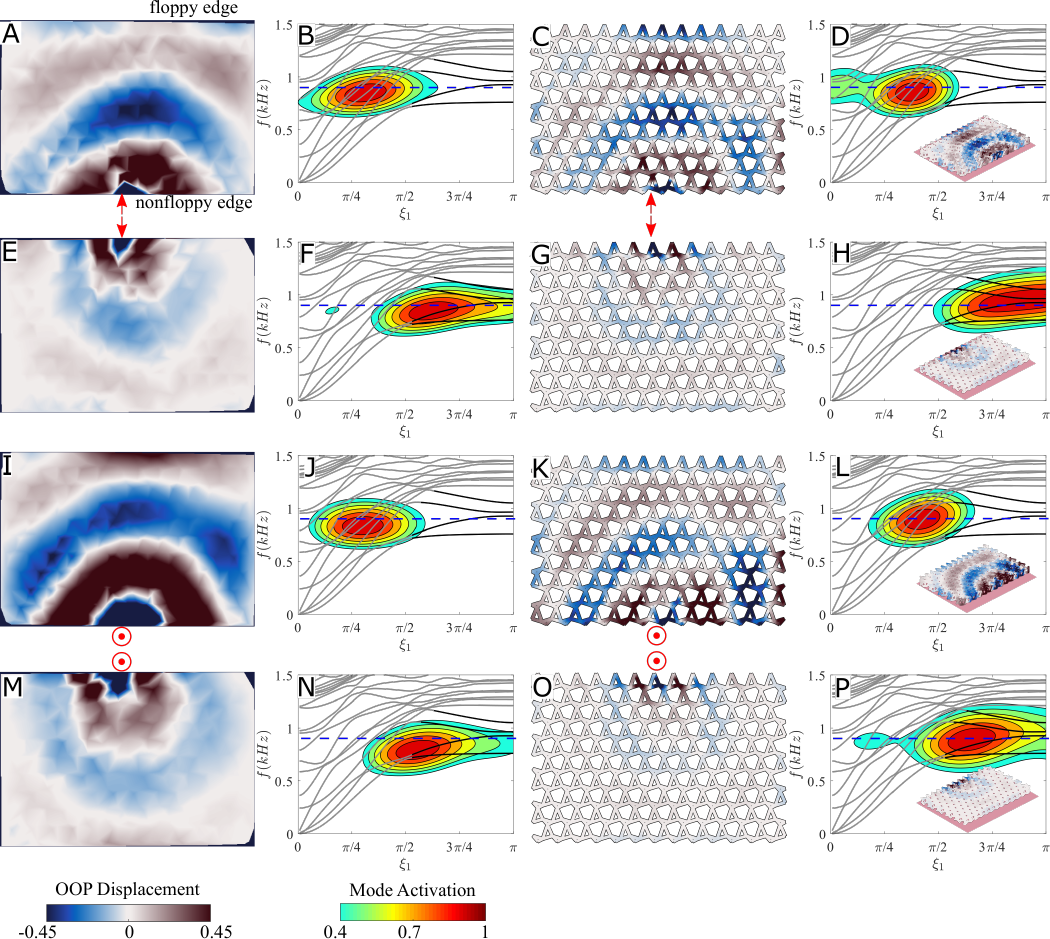}
	\caption{Wavefield and DFT plots for: experiments (A-B) and simulations (C-D) with an in-plane excitation at the rigid edge, experiments (E-F) and simulations (G-H) with an in-plane excitation at the floppy edge, experiments (I-J) and simulations (K-L) with an out-of-plane excitation at the rigid edge, and experiments (M-N) and simulations (O-P) with an out-of-plane excitation at the floppy edge, all carried out at a carrier frequency of $900\, \textrm{Hz}$ (dashed blue lines). Colorbars in panel M and N apply to all wavefields and contours in the figure, respectively, with the former representing the displacement intensity in the wavefield and the latter highlighting modal activation on the excited edge. While the wavefields capture the signature of wave propagation on the bilayer, the insets of D, H, L, and P allow to appreciate propagation through the 3D structure of the bilayer. On the wavefields, in-plane excitation is denoted by the arrows pointing up and down on an edge, and out of plane excitation is denoted by two concentric circles (i.e. arrow pointing out of the page) on an edge. Note that the wavefields and contours are normalized by the highest value in their respective data sets.}\label{fig:res900}
\end{figure*}

We now experimentally demonstrate the activation of the polarized modes in the bilayer prototype shown in Fig. \ref{fig:system}. To this end, we excite the bilayer with, first, in-plane and, later, out-of-plane tone bursts, in the frequency range of the topological edge modes found via supercell analysis, and we perform 3D laser Doppler vibrometer measurements on the joints of the bilayer surface. We verify the proper activation of the desired bulk or edge modes through a morphological inspection of the wavefield data, and a spectral analysis of the transformed wave response in the $\xi_1-\omega$ plane. Furthermore, we ensure that the edge-selective polarization is frequency-selective, thus providing additional evidence that the activated modes are topological in nature. All results are corroborated by full-scale wave propagation simulations of a bilayer geometry, discretized by 3D Timoshenko beams.

The prototype is excited with a 5-cycle burst applied at the tip of the center-most cell of the top and bottom edge, respectively. We excite the structure at a carrier frequency of $900\,\textrm{Hz}$ which lies in the frequency range $760 - 1050\,\textrm{Hz}$, where the topological modes fold and display the highest density of states, to maximize their signatures in the resulting wavefield. The relevant region of the band diagram is shown by the black portions of branches 1-4 in Fig. \ref{fig:beam_bdms}A. 

Results for in-plane excitation with a carrier frequency at $900\, \textrm{Hz}$ prescribed at the bottom edge are shown in Figs. \ref{fig:res900}A and B, and corresponding results for excitation at the top edge are shown in Figs.  \ref{fig:res900}E and F. Snapshots of the resulting out-of-plane wavefields show strong asymmetry between the top and bottom edge excitations: the top edge excitation produces highly localized deformation at that edge, while excitation at the bottom edge produces a bulk-like flexural wave. This corroborates the results from the ideal Maxwell case and the supercell, where the top edge exhibits floppy edge behavior, while the bottom edge is nonfloppy.

We verify activation of the topological modes via spectral analysis of the response. We collect the time histories of the out-of-plane displacement at evenly spaced points along the edge where the excitation is applied, perform a 2D discrete Fourier transform (DFT) on this spatio-temporal data matrix, and superimpose the contours of the resulting spectral amplitude surfaces onto the supercell band diagram to infer which modes are predominantly activated. The DFT analysis corroborates the conclusions made from the inspection of the wavefields, supporting the notion of topological polarization. Excitation at the floppy edge produces a spectral signature that is spread across the branches associated with the edge modes. In contrast, excitation of the nonfloppy edge results in activation of longer wavelength modes (at lower values of $\wavenumber_1$) that belong to the flexural bulk band.

These results are confirmed by simulations performed on a full-scale model of the bilayer, discretized with 3D Timoshenko beam elements (with the same element characteristics used in the supercell analysis). The results for the same excitation conditions used in the experiments (excitation from the nonfloppy edge in Figs. \ref{fig:res900}C-D, and from the floppy edge in Figs. \ref{fig:res900}G-H) match the experiments qualitatively, with only a relatively small deviation in frequency that can be attributed to some inevitable discrepancies in material properties between the model and prototype (due to material property variability resulting from the additive manufacturing process) and to other non-idealities in the geometry of the specimen.

Asymmetric behavior for flexural waves triggered by in-plane excitation is observed in the bilayer presented in Ref.~\cite{charara2021topological}, albeit with a significantly milder polarization signature. However, in that case, such asymmetry requires a direct in-plane strain activation of the polarized layer, and is completely lost when the excitation is prescribed out-of-plane. We repeat the experiments with the current bilayer using an out-of-plane tone burst to test the robustness of the asymmetry achievable against changes in the orientation of the excitation force (i.e., in-plane vs out-of-plane). We show the resulting wavefields and DFTs for excitations from the nonfloppy (Figs. \ref{fig:res900}I and J) and floppy (Figs. \ref{fig:res900}M and N) edges. The results are consistent with their in-plane excitation counterparts: the wavefield snapshots reveal a highly polarized response with displacement localized to the edge when we excite at the floppy edge, and propagation deep into the bulk for excitation at the nonfloppy edge. Again, the DFT plots confirm this dichotomy of the mode activation. The magnitude of this result can be truly appreciated by recalling that the topological character of the bilayer stems from the geometry of its two kagome layers. The dichotomous behavior between the edges documented in Fig. \ref{fig:res900} confirms that the coupling provided by the interlayer connections is very effective in transferring the topological character to the flexural modes, yielding omnimodal polarization encompassing both in-plane and out-of-plane behavior. The results are again confirmed by full-scale numerical simulations (Figs. \ref{fig:res900}K, L, O, and P), with the DFTs showing, even more clearly in this case, topological polarization of the edge modes.

\begin{figure}[t!]
	\centering
	\includegraphics{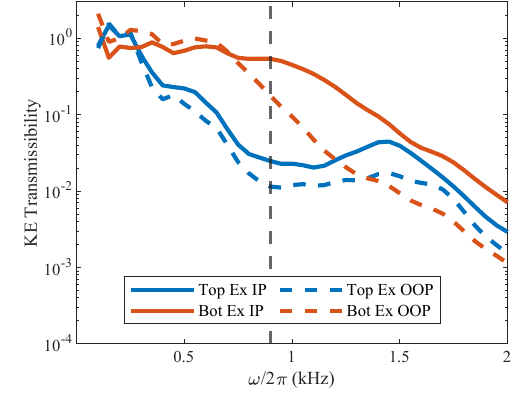}
	\caption{Computational kinetic energy transmissibility of simulated waves at carrier frequencies ranging between $100-2000\textrm{Hz}$, at $50\textrm{Hz}$ intervals, for excitations using IP and OOP forces. Transmissibility is calculated over a time range where we expect reflections within the lattice to be minor, and the data is normalized prior to plotting. The vertical dashed line highlights $900\textrm{Hz}$.}
    \label{fig:transKE}
\end{figure}

\response{
For completeness, we calculate the kinectic energy transmissibility of the lattice over a wide range of frequencies to assess the level of frequency selectivity of the omnimodal polarization achievable in a spectral region containing the polarized modes (Fig. \ref{fig:transKE}). We perform full-scale wave propagation simulations on the Timoshenko beam-discretized lattice model using IP (solid lines) and OOP excitations (dashed lines) applied at the floppy (blue lines) and nonfloppy edges (orange lines). We sweep the frequency of excitation and, for each frequency, we normalize the kinetic energy reaching the output edge by that sampled along the input edge over a time range in which the wave can be considered transient. The polarization, which manifests as edge-selectivity of localization, is conveniently identifiable as the gap between each pair of lines (i.e., solid and dashed). The largest difference in transmissibility happens in the frequency range where the polarized modes are flattest on the band diagram ($\sim900\textrm{Hz}$), for both types of excitation. The persistence of edge-selectivity at frequencies lower and higher than the precise spectral regions of the edge modes is explainable by invoking the following considerations. At lower frequencies, the polarized modes are still present, albeit much longer in wavelength and, thus, competing with nearby bulk-like modes in the same spectral regions of the band diagram, which makes them harder to isolate from the modally-dense response. Moreover, it is worth noting that the tone bursts used in the simulations and experiments are Hanning windowed, resulting in the inevitable activation of a finite frequency band about the carrier. At higher frequencies, such widening of the excitation spectrum is especially pronounced, spreading the wave's energy over a band of frequencies higher and lower than the intended carrier frequency, which results in partial activation of the polarized modes. The pairs of lines eventually coalesce, signifying that symmetric wave behavior is re-established when the frequency of excitation enters far enough into the bulk band, and all the excited phonons are bulk-like in nature. Finally, the consistent drop in transmissibility from all lines as the frequency increases can be explained by the decrease in wave speed with the increase of the activated frequencies, which reduces the kinetic energy effectively reaching the output side within the calculated time frame.
}

\section{Concluding Remarks}
We have introduced a broad family of topological mechanical metamaterials, referred herein as mirror-symmetric Maxwell bilayers, and explored their mechanical response via analytical calculations, numerical simulations, and experimental testing. We have provided an analytical theory of how omnimodal topological polarization arises for coupled in-plane and out-of-plane deformations of ideal Maxwell bilayers in the spring-mass \response{model}. We have also shown that these modes remain localized at finite frequencies \response{in structural Maxwell bilayers}. Finally, we have experimentally verified the existence of this omnimodal polarization via laser vibrometry experiments.

\response{
An important aspect of the geometry studied in the present manuscript is the mirror-symmetry and the particular choice of interlayer connections which leverage the known topological polarization of the in-plane mechanics of Maxwell lattices and provides a design principle for mechanical bilayers with topological polarization in both the in-plane and out-of-plane domains; however, more generic geometries may yield stronger polarizations and could be determined by applying tools such as machine learning to navigate the high-dimensional design space. Furthermore, The connection between the topologically protected zero modes in the spring-mass model and the edge-localized modes at finite frequencies in the beam model warrants further study. It is a special manifestation of a more general class of dilution mechanisms observed in topological mechanics problems in going from ideal to structural lattices, e.g., Ref.~\cite{ma2018edge}. The present manuscript utilizes the aspect ratio of the beams as an effective parameter to distinguish between regions where the low-frequency, short-wavelengths modes localize to a particular edge, but does not explicitly prove that the floppiness is topologically protected. The primary difficulties towards achieving this goal are the fact that the beam model is over-constrained, so that the winding number cannot be computed. The continuum approximation developed in Ref.~\cite{sun2020continuum} could potentially be extended to address this issue to show the topological protection of long-wavelength edge modes in the structural lattice. Simplified models that treat these bonds as filaments with bending and stretching rigidity~\cite{mao2013effective,mao2013elasticity} may also provide insight towards this detail.

The topologically-polarized bilayers explored in this manuscript offer functionalities in ways not achievable by previous mechanical metamaterials. The particular utility of these structures is their topological polarization, which endows the materials with insulating characteristics. Importantly, the topological protection against defects, damage, and wear and tear of the material represents an attribute of significant practical utility for the design of structural lattices intended to bear load and be subjected to dynamic excitations such as impact and vibrations. In addition, this property can yield stress localization at topological domain walls. Such stress focusing has been implemented theoretically in two-dimensional planar lattices~\cite{zhou2018topological} and experimentally in three-dimensional cellular lattices~\cite{paulose2015topological}, and the results presented herein could be utilized to focus stress from flexural deformations of the lattice in a similar manner. This stress focusing could be used to redistribute mass towards the high-stress regions during the design process, thereby enhancing the strength against failure of the structure and the ability to resist shape change. Furthermore, the individual layers possess nonlinear Guest-Hutchinson modes in the spring-mass limit~\cite{guest2003determinacy,rocklin2017transformable}, whose interplay with flexural modes may be used to control the curvature of the bilayer.
}

\begin{acknowledgments}
This work was supported by the National Science Foundation (NSF Grant No. EFRI-1741618 M.C., S.G.) and the Office of Naval Research (MURI N00014-20-1-2479 J.M., K.S., X.M.) and leveraged the High Performance Computing (HPC) systems at the Minnesota Supercomputing Institute (MSI).
\end{acknowledgments}

\bibliography{mainReferences.bib}

\section{Materials and Methods}
\subsection{Zero modes in the spring-mass model}
\response{The zero modes of a spring-mass networks span the nullspace of the compatibility matrix, which can be considered under periodic boundary conditions. For bulk modes, this is achieved by performing the Fourier transformation:

\begin{equation} \label{eq:fourier}
    \compatibility(\wavevector) = \sum_\cell \compatibility(\cell) e^{-i \wavevector \cdot \cell}, 
\end{equation}

\noindent
where the Bloch factor $^{-i \wavevector \cdot \cell}$ is non-trivial only for bonds that connect to adjacent cells. Thus, the assignment of such intercellular connections determines which particular Bloch factors appear in the compatibility matrix and the appropriate choice is the \emph{symmetric gauge} where there are an equal number of factors $e^{i \wavenumber_{1,2}}$ and $e^{-i \wavenumber_{1,2}}$, which eliminates superfluous phase factors in the description of the edge modes~\cite{kane2014topological,rocklin2017directional,sun2020continuum}. For edge modes, the Fourier transform in Eqn.~(\ref{eq:fourier}) is extended to the Laplace transform to allow for complex-valued wavevectors. }

\response{The Maxwell criterion for the lattice implies that each block of the compatibility matrix is square, therefore} its nullspace is non-empty at wavevectors for which its determinant vanishes: $\det \compatibility(\wavevector) = 0$. In general, this determinant is a Laurent polynomial in the Bloch factors, $\sum c_{m_1, m_2} e^{i (m_1 \wavenumber_1 + m_2 \wavenumber_2)}$, where the sum is taken from $m_{1,2} = -3$ to $m_{1,2} = 3$ in the symmetric gauge and $c_{m_1, m_2}$ are real-valued coefficients that depend on the lattice geometry. \response{Thus, for each value of $\wavenumber_1$ in the Brillouin zone, there are up to six complex values of $\wavenumber_2$ for which this determinant vanishes and there exists an edge mode. Consequently, the phase of the determinant of the compatibility matrix changes over any loop enclosing such a zero~\cite{stone2002mathematics}. In particular, the winding number in Eqn.~(\ref{eq:winding}) considers a contour over the bulk modes $\decay_2 = 0$, which bound the modes localized to the top edge, $\decay_2 < 0$, and those localized to the bottom edge, $\decay_2 > 0$. The gauge symmetry implies that a winding number of $\winding=0$ corresponds to three modes on each edge, whereas a winding number of $\winding = -3$ corresponds to six modes on the top edge. Finally, this contour cannot be computed when it crosses a zero, as is the case in the presence of bulk zero modes.}

\subsection{Lattice Details}

The lattice is 3D printed using SLS~\cite{kumar2003selective} (DTM Sinterstation) by Stratasys Direct. The lattice is first modeled using the Solidworks software and exported  as a ``.STL'' file at a resolution finer than that of the SLS process used (~$0.76\, \textrm{mm}$ in $x-y$, and ~$0.1\, \textrm{mm}$ in $z$). SLS consists of depositing a fine layer of powder material (Stratasys Nylon 12-glass-fiber-reinforced nylon -- Young's Modulus = 2.896 GPa and Poisson's ration = 0.35) which is then melted by a laser into a solid layer following the cross section of a given design, bonding it to all previously deposited layers. This printing process produces a monolithic structure, while the more commonly commercially available fused deposition modeling (FDM) often results in imperfect bonding between deposited layers and adjacent deposited lines that can detrimentally affect wave transport properties. These considerations were key factors in the selection of SLS as the fabrication method for the current task.

The slenderness ratio and cell size are decided concurrently, making sure that the thickness of the beams is within the resolution of the SLS 3D printing process and that the beams are sufficiently slender to prevent their flexural behavior from completely overwhelming the lattice response. The lattice dimensions are selected to be large enough to appreciate the topologically induced decay phenomenon, within the dimensions of the 3D printer's fabrication print bed ($67.31\times34.29\,\textrm{cm}$).

\subsection{Computational Beam Model}
\response{
The bilayer connections are modeled as 3D Timoshenko beams \cite{karadeniz2013finite}. Recall that that each lattice bond is discretized into multiple beam elements, featuring three positional degrees of freedom and three cross sectional rotations. The stiffness matrix of a single beam element is given by $\dynamical_b = \int_L \compatibility_b^T \mathbf{k} \compatibility_b~ dL$, where $L$ is the length of the beam element, $\compatibility_b$ is a matrix of derivatives of the shape functions, which approximate the compatibility relations within one element, and $\mathbf{k}$ is a diagonal constitutive matrix of elastic constants, featuring rigidity contributions for axial deformation (scaling with $\radius^2$), lateral deflection (scaling with $\radius^4$) [encompassing the effects of, both, shear and bending deformability encoded in the beam model], and torsional deformation (scaling with $\radius^4$). The shape functions used in the calculation of the elemental stiffness matrix $\dynamical_b$ are linear in axial and torsional deformations, cubic in flexural deformation, and quadratic in rotations. Stiffness matrices of each individual beam are assembled into a global stiffness matrix for the entire system, which is used for computational analysis. Note that, for each discretized beam element, $\compatibility_b$ plays a role analogous to a single row of the compatibility matrix in the spring-mass counterpart, which quantifies the constraints for a single spring. Further details of the framework used in this work can be found in the SI Appendix and a full, pedagogical account can be found in Ref.~\cite{karadeniz2013finite}.}

We discretized each beam into 9 to 11 elements, depending on the bond length (sufficient discretization to allow for appropriate mode visualization), resulting in a model with 3240 and 29376 nodes for the supercell and full-scale lattice models, respectively. For supercell calculations of $100$ unit cells, the beams were discretized into only 5 elements due to computational constraints; however, comparing the results of the higher and lower discretization reveals a change of less than 2\% for all the relevant modes (lowest 6), meaning that this alternative discretization is accurate in capturing the spectral characteristics of the modes, despite offering a lower spatial resolution of the mode shapes.

\response{
\subsection{Supercell construction and band diagram calculation}

To calculate the band diagram of the supercell, we consider a supercell consisting of 10 cells, discretized using 3D Timoshenko beam elements, and connected in the $\latticevector_2$ direction with free boundary conditions at the ends, and Bloch-periodic boundary conditions with respect to wavenumber $\wavenumber_1$ in the $\latticevector_1$ direction.~
We extract the Bloch-reduced arrays of generalized nodal displacements  $\tilde{\Displacements}$ and cross-sectional rotations $\tilde{\Rotation}$ in the supercell as the modes of $\tilde{\dynamical}_b(\wavenumber_1)\left[\tilde{\Displacements}~ \tilde{\Rotation}\right]^T= \omega^2 \tilde{\mass}_b(\wavenumber_1) \left[\tilde{\Displacements}~ \tilde{\Rotation}\right]^T$, where $\tilde{\dynamical}_b(\wavenumber_1)$ [$\tilde{\mass}_b(\wavenumber_1)$] is the stiffness [mass] matrix of the supercell under Bloch conditions.

}

\subsection*{3D Laser Doppler Vibrometer Experiments}
\begin{figure}[t]
	\centering
	\includegraphics{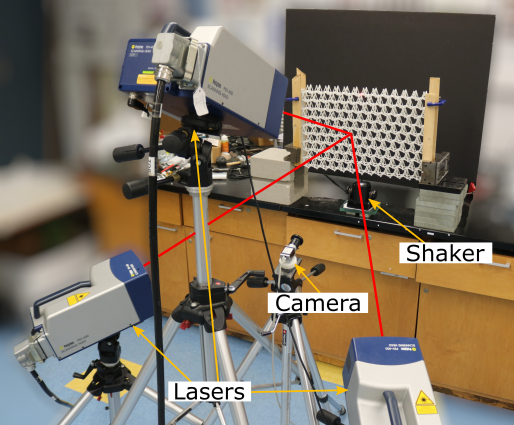}
	\caption{Experimental setup for the 3D scanning laser Doppler vibrometry (SLDV) testing of the 3D printed bilayer lattice. }\label{fig:expsetup}
\end{figure}

The experimental setup is shown in Fig. \ref{fig:expsetup}. A 3D laser Doppler vibrometer (Polytec PSV-400-3D) is used to measure the velocity of the joints of the beams (the selected scan sites) on the face of the lattice closest to the lasers, where reflective tape is applied to increase the material reflectivity, in order to reduce noise in the data. An electromechanical shaker (Br\"{u}el \& Kj\ae r Type 4810), internally triggered by the vibrometry setup through an amplifier (Br\"{u}el \& Kj\ae r Type 2718), probes the lattice through a stinger at the desired excitation location, and a 5-cycle Hann-windowed burst, with a carrier frequency of $300, 900,~\textnormal{or}~ 1300\, \textrm{Hz}$ is fired into the structure while velocity is measured by the laser heads at a scan point. 
The process is repeated for each scan point, automatically moving the lasers to subsequent scan locations in a prescribed sequence, and providing enough relaxation time between scans to ensure bursts have fully dissipated by damping before the next measurement is taken.

The velocity data collected is decomposed into $\hat{x},\hat{y}$ and $\hat{z}$ components using the Euler angles internally calculated by the vibrometry software. This data is further processed in MATLAB to recreate the wave fields and DFT plots. The temporal data of the spaces between the scan points is interpolated by MATLAB post-processing using a triangulation-based linear interpolation algorithm. It should be noted that this interpolation favors the creation of bulk-appearing wave fronts, which implies that the wave fields for experiments with excitation at the floppy side could potentially be experiencing higher localization than what is shown in the result figures.

\appendix

\onecolumngrid

\response{
\section{Discussion of notation}

The main text studies topological polarization in the context of dynamics, and hence requires bridging two disparate notation systems: one from the physics community and one from the mechanics community. This is achieved by utilizing the mechanics notation wherever possible. This appendix discusses the distinctions between the two notations.

In the physics literature, when we have a mechanical systems composed of point masses connected by harmonic springs, the matrix capturing the kinetic interaction between the masses is referred to as the dynamical matrix, typically represented by the symbol $\mathbf{D}$. This matrix maps the displacements of the lattice sites to the forces balancing at the sites, $\mathbf{F} = \mathbf{D} \mathbf{U}$, where $\mathbf{F}$ is an array of forces and $\mathbf{U}$ is an array of displacements). In the mechanics literature, this object is more typically called a stiffness matrix, typically represented by the symbol $\mathbf{K}$. Moreover, in certain dynamics contexts, the dynamical matrix refers to the quantity $\mathbf{D}=-\omega^2\mathbf{M}+\mathbf{K}$, which is the matrix that links forces and displacements in the frequency domain (Fourier-transformed equations of motion) for a dynamical problem. This discrepancy is reinforced by the fact that the matrix of spring constants (i.e., the coefficients of Hooke's law) are written $\mathbf{K}$ in the physics literature. This typically corresponds to the constitutive matrix, written as $\mathbf{C}$, in the mechanics literature which is not to be confused with the symbol for the compatibility matrix that maps from vertex displacements to bond extensions, $\mathbf{E} = \mathbf{C} \mathbf{U}$, in the physics literature. Thus, this compatibility matrix is instead written as $\mathbf{B}$, as used in mechanics, and the spring constants are written as $E$ to signify their connection to the Young's modulus of beam elements. Finally, the wavenumbers in the physics literature are typically written as $q$ or $k$, whereas the main text uses $\xi$ to accomodate the mechanics notation.
}
\response{
\section{Topological polarization of planar kagome lattices}

The main text leverages analytical results for the topological polarization of planar kagome lattices to determine the omnimodal topological polarization of  mirror-symmetric kagome bilayers. This appendix reviews the known results for the planar kagome lattice.

The planar kagome lattice consists of cells that contain three vertices and six edges so that it is critically-coordinated in the bulk. These vertices are labeled in Fig.~\ref{fig:sifig1}A as $\position_1$, $\position_2$, and $\position_3$ along with the two lattice vectors $\latticevector_1$ and $\latticevector_2$. The six total edges of the unit cell consist of the three that connect the vertices \emph{within} the cell to one another and edges that connect the vertices to adjacent cells: $\position_2(0,0)$ connects to $\position_1(1,0)$, $\position_3(0,0)$ connects to $\position_2(-1,1)$, and $\position_1(0,0)$ connects to $\position_2(0,-1)$. These particular intercellular connections, in contrast to the dashed lines shown in Fig.~\ref{fig:sifig1}A which belong to adjacent cells, are chosen to achieve a symmetric gauge.

The corresponding Bloch-periodic compatibility matrix has six rows and six columns, and its determinant is a second-order polynomial in the two Bloch factors $z_{1,2} = e^{i \wavenumber_{1,2}}$. Thus, the planar kagome lattice generically exhibits two topological edge modes, which can be understood by the fact that two edges (constraints) are removed from each cell on the boundary in a finite-size kagome geometry. The localization of these zero modes is shown via bulk analysis by the strictly negative decay rates, $\decay_2 \leq 0$, along the $\latticevector_2$-direction, in Fig.~\ref{fig:sifig1}B and the constant winding number, $\winding_2$, computed by integrating over the wavenumber, $\wavenumber_2$, in the $\latticevector_2$-direction from $-\pi$ to $\pi$, in Fig.~\ref{fig:sifig1}C for all values of the wavenumber, $\wavenumber_1$, in the $\latticevector_1$-direction. Importantly, the topological phase space of kagome lattices can be characterized analytically, as shown in Ref.~\cite{kane2014topological}. 

\begin{figure}
    \centering
    \includegraphics{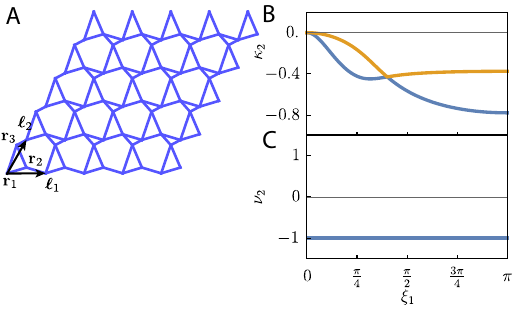}
    \caption{Topological polarization of the planar kagome lattice. (A) View of the lattice and specification of the unit cell with vertex positions $\position_1 = (0,0)$, $\position_2 = (1,0.3)$, $\position_3 = (0.5, \sqrt{3}/2 + 0.6)$, and lattice vectors $\latticevector_1 = (2,0)$, $\latticevector_2 = (1,\sqrt{3})$. The dashed lines signify edges that belong to adjacent cells. (B) The decay rate of zero modes, $\decay_2$, along the $\latticevector_2$-direction as a function of the wavenumber, $\wavenumber_1$, in the $\latticevector_1$-direction. (C) The winding number, $\winding_2$, computed by integrating over the wavenumber, $\wavenumber_2$, in the $\latticevector_2$-direction from $-\pi$ to $\pi$ as a function of the wavenumber, $\wavenumber_1$, in the $\latticevector_1$-direction.}
    \label{fig:sifig1}
\end{figure}
}
\section{Flat bands in coplanar Maxwell bilayers}

The main text presents mirror-symmetric kagome bilayers with a choice of interlayer connections that exhibit lines of bulk modes in the absence of height modulations within the layers. Here, we present an alternative choice of interlayer connections that instead yields surfaces of bulk modes known as \emph{flat bands}.

\begin{figure}
    \centering
    \includegraphics[width=0.5\textwidth]{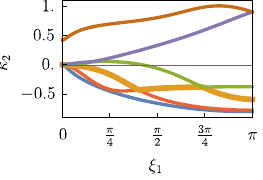}
    \caption{Decay rates for a non-coplanar bilayer whose interlayer connections exhibit a flat band in the coplanar limit, which inhibits prediction of where the two remaining zero modes localize.}
    \label{fig:flatband}
\end{figure}

The key distinction between coplanar bilayers with lines of bulk zero modes and those with surfaces of bulk zero modes is that the interlayer connections in the former constrain an infinite number of vertices whereas in the latter they constrain a finite number of vertices. For the kagome bilayer, this second condition arises for the following interlayer connections: $\position_1^- (0,0)$ connects to $\position_2^+ (0,0)$ connects to $\position_3^-(1,-1)$ connects to $\position_1^+(1,-1)$ connects to $\position_2^-(0,-1)$ connects to $\position_3^+(0,-1)$ connects to $\position_1^-(0,0)$. Thus, the vertices can simultaneously displace out-of-plane at any wavevector provided that their displacements are identical to one another:
\response{
\begin{equation}
    \oop_1^- = \oop_2^+ = \oop_3^- e^{-i (\wavenumber_1 - \wavenumber_2)} = \oop_1^+ e^{-i (\wavenumber_1 - \wavenumber_2)} = \oop_2^- e^{i \wavenumber_2} = \oop_3^+ e^{i \wavenumber_2}.
\end{equation}
}
\noindent
Since such zero modes exist at arbitrary wavevectors, the lowest band is entirely flat. Importantly, the introduction of small height modulations to make the bilayer noncoplanar yields decay rates which cannot be described as a perturbation to the flat band, as shown in Fig.~\ref{fig:flatband}, which makes it more difficult to use this type of connectivitiy to achieve fully polarized Maxwell bilayers.

\section{Computational Beam Model: Details}

In this work, we choose to work with Timoshenko beams to ensure that the results are robust even for short connections whose slenderness ratio may exceed the bounds for which Euler-Bernoulli beam theory is acceptable.

In 3D, the stiffness matrix for a single beam, $\dynamical_b$, encompasses stiffness contributions associated with axial, shear, bending, and torsional deformation. This is captured by reordering rows and columns and partitioning $\dynamical_b$ into minors that highlight the contributions of different mechanisms. Recalling that $\dynamical_b$ links an array $\mathbf{f}_m$ of nodal forces and moments, to an array $\left[ \displacement_m~ \rotation_m \right]^T$ of nodal displacements and rotations, the matrix partition can be written as

\begin{equation} \label{eq:Dbeam}
    \dynamical_{b_{12\times12}} = \begin{pmatrix} 
        \dynamical_{A_{2\times2}} & \mathbf{0} & \mathbf{0} \\
        \mathbf{0} & \dynamical_{S_{4\times4}} & \dynamical_{M_{4\times6}} \\
        \mathbf{0} & \dynamical_{M_{4\times6}}^T &\dynamical_{B_{6\times6}}
    \end{pmatrix}.
\end{equation}

\noindent
where $\dynamical_A$ is the contribution governing axial stiffness, $\dynamical_S$ and $\dynamical_B$ are square matrices controlling shear, and torsion and bending, respectively, and $\dynamical_M$ is a rectangular mixed matrix coupling shear and bending:
\response{
\begin{equation} \label{eq:Dbeam}
    \dynamical_{A} = \begin{pmatrix} 
        \frac{\youngsmod A}{L} & -\frac{\youngsmod A}{L} \\
        -\frac{\youngsmod A}{L} & \frac{\youngsmod A}{L}  \\
    \end{pmatrix}
\end{equation}
\begin{equation}
    \dynamical_{S} = \begin{pmatrix} 
        \frac{12\youngsmod \momOfI_z \shearCorr_y}{L^3} & 0 & -\frac{12\youngsmod \momOfI_z \shearCorr_y}{L^3} & 0\\
        0 & \frac{12\youngsmod \momOfI_y \shearCorr_z}{L^3} & 0 & -\frac{12\youngsmod \momOfI_y \shearCorr_z}{L^3}\\
        -\frac{12\youngsmod \momOfI_z \shearCorr_y}{L^3} & 0 & \frac{12\youngsmod \momOfI_z \shearCorr_y}{L^3} & 0 \\
        0 & -\frac{12\youngsmod \momOfI_y \shearCorr_z}{L^3} &0 & \frac{12\youngsmod \momOfI_y \shearCorr_z}{L^3}
    \end{pmatrix}
\end{equation}.
\begin{equation}
    \dynamical_B = \begin{pmatrix} 
        \frac{\shearMod \polarMom}{L}   & 0     & 0     & -\frac{\shearMod \polarMom}{L}    & 0     & 0 \\
        0       & \frac{\youngsmod\momOfI_y (3\shearCorr_z+1)}{L}   & 0 & 0 & \frac{\youngsmod\momOfI_y (3\shearCorr_z-1)}{L} & 0 \\
        0   & 0     &\frac{\youngsmod\momOfI_z (3\shearCorr_y+1)}{L} & 0 & 0 & \frac{\youngsmod\momOfI_z (3\shearCorr_y-1)}{L}\\
        -\frac{\shearMod \polarMom}{L} & 0 & 0 & \frac{\shearMod \polarMom}{L} & 0 & 0  \\
        0  & \frac{\youngsmod\momOfI_y (3\shearCorr_z-1)}{L} & 0 &0 & \frac{\youngsmod\momOfI_y (3\shearCorr_z+1)}{L} & 0 \\
         0 &0 & \frac{\youngsmod\momOfI_z (3\shearCorr_y-1)}{L}& 0 & 0 &\frac{\youngsmod\momOfI_z (3\shearCorr_y+1)}{L} 
    \end{pmatrix}
\end{equation}.
\begin{equation}
    \dynamical_M = \begin{pmatrix} 
       0 & 0 & \frac{6 \youngsmod \momOfI_z \shearCorr_y}{L^2} & 0 & 0 & \frac{6 \youngsmod \momOfI_z \shearCorr_y}{L^2}\\
       0 & -\frac{6 \youngsmod \momOfI_y \shearCorr_z}{L^2} & 0 & 0 &  -\frac{6 \youngsmod \momOfI_y \shearCorr_z}{L^2}&0 \\
       0 & 0 & -\frac{6 \youngsmod \momOfI_z \shearCorr_y}{L^2} & 0 & 0 & -\frac{6 \youngsmod \momOfI_z \shearCorr_y}{L^2}\\
       0 & \frac{6 \youngsmod \momOfI_y \shearCorr_z}{L^2} & 0 & 0 &  \frac{6 \youngsmod \momOfI_y \shearCorr_z}{L^2} &0
    \end{pmatrix}
\end{equation}
}
\noindent
where $\youngsmod$ is the Young's modulus, $A=\pi R^2$ is the cross-sectional area of the beam, L is the length of the beam element, $I$ is the second moment of area of the cross section about the $\hat{y} ~\textnormal{and}~ \hat{z}$ axes ($I=\pi \radius^4/4$ with $\radius$ the cross section radius), $\shearCorr$ is the shear correction factor for Timoshenko beams \cite{karadeniz2013finite}, $\shearMod$ is the shear modulus, and $J$ is the polar second moment of area of the cross section ($J=\pi \radius^4/2$), is the Young's modulus, A is the cross-sectional area of the beam.  Note that because the beam element matrices are calculated in a local coordinate system where the $\hat{x}$ axis is along the length of the beam, they must be rotated upon assembly into a global system matrix.

As previously mentioned, the stiffness matrix of a beam element is calculated by carrying out the integral $\dynamical_b = \int_L \compatibility_b^T \mathbf{k} \compatibility_b dL$ over the length of the element. $\mathbf{k}$ is diagonal constitutive matrix of elastic coefficients relating an array $\stress$ of stresses and moments to an array $\strain$ of strains, shears, and curvatures (i.e. $\stress = \mathbf{k} \strain$) and is given by

\begin{equation} \label{eq:rigidity}
    \mathbf{k} = \begin{pmatrix} 
        EA & 0      & 0     & 0  & 0  & 0 \\
        0  & GAk    & 0     & 0  & 0  & 0 \\
        0  & 0      & GAk   & 0  & 0  & 0 \\
        0  & 0      & 0     & GJ & 0  & 0 \\
        0  & 0      & 0     & 0  & EI & 0 \\
        0  & 0      & 0     & 0  & 0  & EI  \\
    \end{pmatrix}.
\end{equation}

\noindent
where $k$ is the area shear correction factor. This matrix features elastic constants for axial deformation (row 1), shear deformation (rows 2 and 3), torsion (row 4), and bending (rows 5 and 6). $\compatibility_b$ is an elemental matrix which captures a similar role in a beam element as the compatibility matrix does in a spring-and-mass bond. $\compatibility_b$ contains derivatives of shape functions [that interpolate the elemental displacements and rotations between nodes]. These shape functions are linear in axial and torsional deformation, cubic in flexural deformation, and quadratic in rotations.

As we change the radius of the beam's cross section, the elastic coefficients for axial and flexural deformation scale at different rates -- the ratio of the former to the latter is $\propto 1/\radius^2$. Thus, even though we never lose the inherent effects of clamped boundary conditions and the storage of bending energy of the beams, as we reduce the cross sectional radius $\radius$, we \emph{asymptotically} approach the dominance of axial deformability typical of the spring-mass case.

In the Timoshenko beam framework, the mass matrix of the beam elements is a non-diagonal \response{consistent matrix}, coupling degrees of freedom in the same manner as $\dynamical_b$. This matrix is calculated by integrating the \response{same} elemental shape functions \response{as $\dynamical_b$}, multiplied by material and cross-sectional properties, over the length of the element. The resulting matrix includes contribution from axial, lateral, rotatory (associated with the tilt of the beam cross-sections), and polar (twist about the beam axis) inertial effects. A detailed account of this derivation, as well as that for the beam's stiffness matrix, is provided in Ref.~\cite{karadeniz2013finite}.

\response{
\section{Wave Propagation: Experiments and Full-Scale Simulations at frequencies away from the polarized modes}

We visualize the frequency selectivity of the polarized behavior by performing experiments and simulations at carrier frequencies away from the floppy modes at $\wavenumber=\pi$. Results are shown in Figs. \ref{fig:res300_AP} and \ref{fig:res1300_AP} for simulations and experiments carried out at carrier frequencies of $300$ and $1300\, \textrm{Hz}$, respectively.  Excitations at $300\, \textrm{Hz}$ reveal nearly identical wavefields when we switch the excitation edge from floppy to nonfloppy, with bulk-like characteristics activated from either side. Although the DFTs from the floppy edge do show some activation of the mode branches endowed with polarized character, the activation occurs at much longer wavelengths, where the decay rate of these modes is negligible (making them resemble the bulk modes), overall resulting in bulk-like behavior dominating the response. Experiments performed with a carrier frequency at $1300\, \textrm{Hz}$ also show largely symmetric behavior, although we still see the persistence of some localization when exciting from the floppy edge. This is likely due to the fact that the excitation energy is spread over a band of frequencies due to the windowing applied to the burst, which causes the edge of the main lobe and the side lobes to excite modes in the topological region.
}

\begin{figure}
	\centering
	\includegraphics{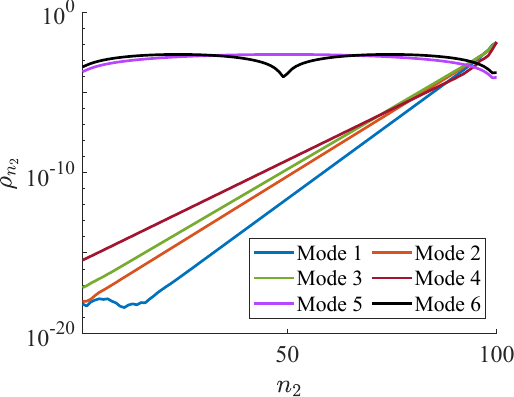}
	\caption{Displacement vs cell index for 100-cell beam-discretized supercell, showing that, even at much longer supercell lengths, beams with the AR used in the lattice for this work do not produce six modes exhibiting floppy edge localization. Unlike the low-$AR$ or spring and mass counterparts, modes 5 and 6 remain bulk-like when we increase the supercell to much longer length, revealing that these are not just long decay length modes that simply appear bulk-like due to the relatively short size of the $10$-cell supercell. Thus, their residence in the bulk band, as shown in the band diagram, is confirmed.} \label{fig:beams100uc}
\end{figure}

\begin{figure*}
	\centering
	\includegraphics{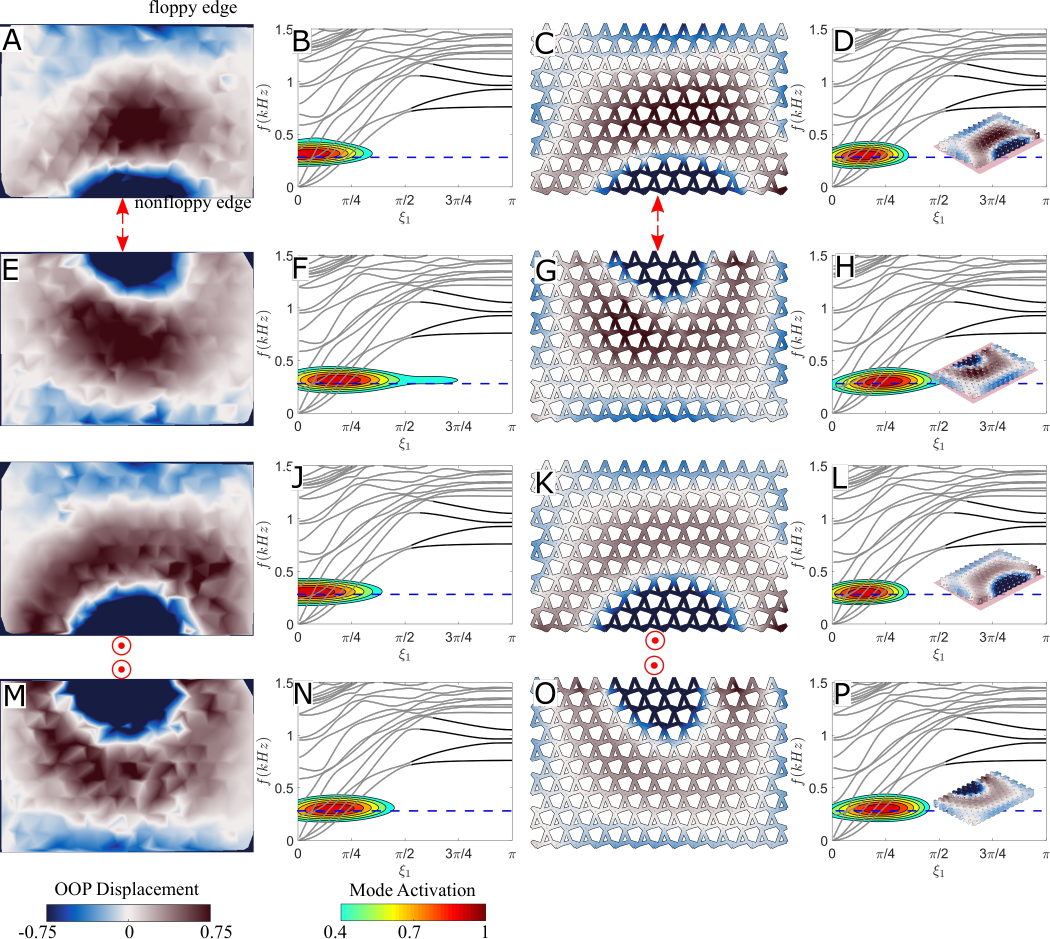}
	\caption{Wavefield and DFT plots for: experiments (A-B) and simulations (C-D) with an in-plane excitation at the rigid edge, experiments (E-F) and simulations (G-H) with an in-plane excitation at the floppy edge, experiments (I-J) and simulations (K-L) with an out-of-plane excitation at the rigid edge, and experiments (M-N) and simulations (O-P) with an out-of-plane excitation at the floppy edge, all carried out at a carrier frequency of $300\, \textrm{Hz}$ (dashed blue lines). Colorbars in panel M and N apply to all wavefields and contours in the figure, respectively, with the former representing the displacement intensity in the wavefield and the latter highlighting modal activation on the excited edge. While the wavefields capture the signature of wave propagation on the lattice surface, the insets of D, H, L, and P allow to appreciate propagation through the 3D structure of the bilayer. Note that the wavefields and contours are normalized by the highest value in their respective data sets. While the main text includes experimental figures highlighting edge-selectivity in the topologically polarized lattice, this image highlights frequency selectivity, an equally important ingredient in ensuring this behavior is not spectrally ubiquitous, providing evidence for the topological nature of the activated modes.}\label{fig:res300_AP}
\end{figure*}

\begin{figure*}
	\centering
	\includegraphics{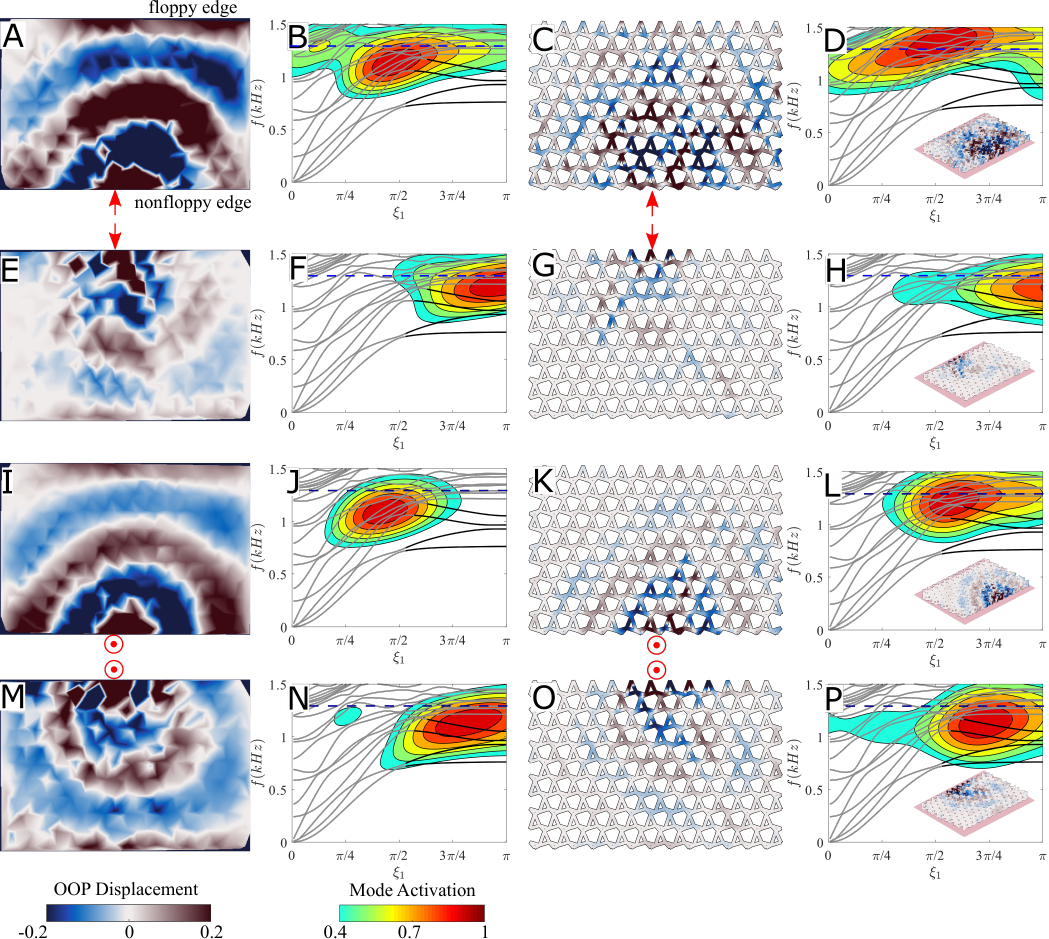}
	\caption{Wavefield and DFT plots for: experiments (A-B) and simulations (C-D) with an in-plane excitation at the rigid edge, experiments (E-F) and simulations (G-H) with an in-plane excitation at the floppy edge, experiments (I-J) and simulations (K-L) with an out-of-plane excitation at the rigid edge, and experiments (M-N) and simulations (O-P) with an out-of-plane excitation at the floppy edge, all carried out at a carrier frequency of $1300\, \textrm{Hz}$ (dashed blue lines). Colorbars in panel M and N apply to all wavefields and contours in the figure, respectively, with the former representing the displacement intensity in the wavefield and the latter highlighting modal activation on the excited edge. While the wavefields capture the signature of wave propagation on the lattice surface, the insets of D, H, L, and P allow to appreciate propagation through the 3D structure of the bilayer. Note that the wavefields and contours are normalized by the highest value in their respective data sets. While the main text includes experimental figures highlighting edge-selectivity in the topologically polarized lattice, this image highlights frequency selectivity, an equally important ingredient in ensuring this behavior is not spectrally ubiquitous, providing evidence for the topological nature of the activated modes.}\label{fig:res1300_AP}
\end{figure*}

\end{document}